# Observational evidence for temporary planetary wave forcing of the MLT during fall equinox

Nora H. Stray[1], Rosmarie J. de Wit[1], Patrick J. Espy[1,2], and Robert E. Hibbins[1,2]

[1]Department of Physics, NTNU, Trondheim, Norway, [2]Birkeland Centre for Space Science, Bergen, Norway

**Abstract** We present direct observations of zonal wave numbers 1 and 2 planetary wave activity in the mesopause region derived from a longitudinal chain of high-latitude Northern Hemisphere (51–66°N) Super Dual Auroral Radar Network radars. Over a 9 year period (2000–2008), the planetary wave activity observed shows a consistent increase around the fall equinox. This is shown to be coincident with a minimum in the magnitude of the stratospheric winds and consequently a minimum in the stratospheric gravity wave filtering and the subsequent momentum deposition in the mesopause region. Despite this, the observed meridional winds are shown to be perturbed poleward and mesopause temperatures rise temporarily, suggesting that westward momentum deposition from planetary waves temporarily becomes the dominant forcing on the mesopause region each fall equinox.

## 1. Introduction

The mesosphere and lower thermosphere (MLT) is driven far from radiative equilibrium by the deposition of momentum by waves [*Fritts and Alexander*, 2003]. Generally, this forcing is due to gravity waves (GWs), generated in all directions in the lower atmosphere, but filtered by stratospheric background winds as they propagate upward. As GWs propagate into the MLT and break, they provide a net momentum source to the region [e.g., *Lindzen*, 1981; *Holton*, 1982, 1983; *Garcia and Solomon*, 1985] and force the residual meridional circulation. During equinox, when the zonal stratospheric winds approach zero, little net gravity wave filtering takes place, and subsequently, little net momentum is deposited into the mesosphere by GWs. Hence, the atmosphere should approach radiative equilibrium in the MLT around equinox [*Andrews*, 2010].

In this paper we present observations of enhanced planetary wave (PW) activity in the MLT during fall equinox that occurs concurrently with increased poleward meridional flow and reduced gravity wave forcing. Mean meridional winds as well as the zonal wave numbers 1 and 2 PW structures in the Northern Hemisphere MLT have been extracted using a longitudinal chain of Super Dual Auroral Radar Network (SuperDARN) radars. These are combined with winds from the UK Meteorological Office (UKMO) Stratospheric Assimilated Data [*Swinbank et al.*, 2013] to investigate the relationship between the PW activity in the MLT and the underlying wind field. A momentum flux meteor radar is also used to assess the forcing of the MLT winds by gravity wave momentum deposition during the fall equinox [*Hocking et al.*, 2001; *Hocking*, 2005; *Fritts et al.*, 2010]. Although some of these features have been previously modeled or observed in isolation, here a suite of observations is presented and combined in order to examine the primary forcing mechanism of the MLT residual circulation and its contribution to the mesospheric temperature enhancements that have previously been reported [*Taylor et al.*, 2001; *Espy and Stegman*, 2002; *French and Burns*, 2004].

## 2. Data

PW activity with zonal wave numbers 1 and 2 for years 2000–2008 was extracted from daily mesospheric meridional meteor wind anomalies using a longitudinal chain (150°W–25°E) of SuperDARN radars [*Greenwald et al.*, 1985, 1995] spanning latitudes from 51 to 66°N as described by *Kleinknecht et al.* [2014]. Tidal effects and effects of the 2 day wave were removed using the techniques outlined in *Hibbins and Jarvis* [2008] and *Kleinknecht et al.* [2014] before the retrieval of the PW structures. Figure 1 shows Hovmöller diagrams of wave numbers 1 and 2 PW components for the year 2000 as an example. Blue and red colors signify equatorward and poleward wind, respectively. Figures for all other years can be found in *Kleinknecht et al.* [2014, Figure 5].

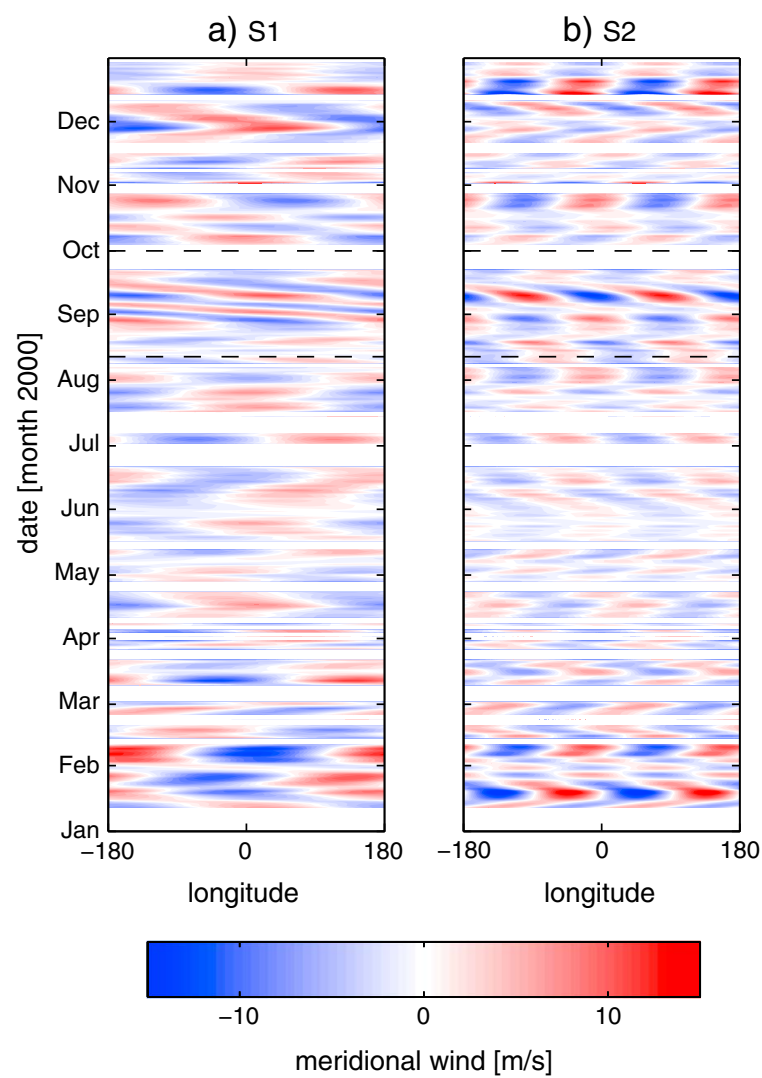

**Figure 1.** Hovmöller diagrams of zonal wave numbers 1 and 2 ($S_1$, $S_2$) of the mesospheric (mean altitude around 95 km) meridional wind, year 2000. Blue and red colors signify poleward and equatorward wind, respectively. A period of westward propagating wave activity can clearly be seen around fall equinox in the $S_1$ mode. Adapted from *Kleinknecht et al.* [2014].

The UK Meteorological Office (UKMO) Stratospheric Assimilated Data winds [*Swinbank et al.*, 2013] from the years 2000–2008 were used to quantify the tropospheric and stratospheric wind regimes. UKMO wind data are available for pressure levels between 1000 and 0.1 hPa (0.3 hPa before 28 October 2003 and 0.4 hPa between 14 March 2006 and 14 May 2007) and a horizontal resolution of 2.5° × 3.75° (latitude × longitude).

Vertical profiles of the density-weighted GW momentum fluxes of zonal momentum, $\overline{\rho u' w'}$, have been determined using a new generation All-Sky Interferometric (SKiYMET) meteor radar [*Hocking et al.*, 2001; *Fritts et al.*, 2010] located in Trondheim (63°N, 10°E), Norway [*de Wit et al.*, 2014]. The system, operating at a frequency of 34.21 MHz, is optimized to measure momentum fluxes by the use of a circular transmitter array consisting of eight antennas, which directs most of the 30 kW peak power between zenith angles of 15° and 50°. Maximum meteor count rates are observed around 90 km, and daily count rates range between 6000 and 12,000 unambiguous meteors at zenith angles between 15° and 50° and altitudes of 70–100 km. Momentum fluxes are derived in four altitude ranges (80–85 km, 85–90 km, 90–95 km, and 95–100 km) after removing the background winds [*Andrioli et al.*, 2013] using the method proposed by *Hocking* [2005] to estimate their vertical divergence using densities, $\rho$, taken from the COSPAR International Reference Atmosphere 1986 monthly mean density data set [*Fleming et al.*, 1990].

## 3. Results

Figure 2 shows the observed mesospheric PW activity (the sum of zonal wave numbers 1 and 2) at high northern latitudes (51–66°N) between 1 March and 15 October for the years 2000–2008. The PW activity is also shown smoothed over 10 days for better visualization. These data are representative of the PW activity around an altitude of ∼ 95 km [e.g., *Hall et al.*, 1997; *Hibbins and Jarvis*, 2008]. The maximum eastward (red) and westward (green) zonal mean zonal wind speed in the height column from 1000 to 0.1 hPa (ground to ∼ 65 km) and at the latitude band between 51 and 66°N are displayed for the same time periods. These winds are extracted from the zonal mean zonal UKMO winds.

Increased PW activity can be observed after spring equinox, during midsummer and around fall equinox. The enhancements in spring and summer show strong interannual variability in both timing and strength. The spring enhancements are related to the final breakdowns of the polar vortex, as discussed by *Manson et al.* [2002] and *Shepherd et al.* [2002] who found good agreement between (final) stratospheric warming events and springtime temperature perturbations. The midsummer peak in PW activity is consistent with interhemispheric propagation of PW's into the summer mesosphere [e.g., *Forbes et al.*, 1995; *Espy et al.*, 1997; *Hibbins et al.*, 2007; *Ford et al.*, 2009; *Hibbins et al.*, 2009] or in situ generated waves [*Nielsen et al.*, 2010]. A detailed description of this feature will be the subject of a subsequent publication. Although there is some interannual variability in both the start date and peak amplitude, there is a persistent PW enhancement that occurs around fall equinox (∼days 230–270) in all years where data are available. To extract seasonal features from the year-to-year variability, the PW amplitude and the maximum wind values between 1 March and 15 October have been averaged over all 9 years. This climatology of PW activity is shown in Figure 3 and shows the enhancement around fall equinox to be a climatologically consistent feature. It is also clear that the observed enhancement in high-latitudinal mesosphere PW activity (the sum of zonal wave numbers 1 and 2) in fall occurs simultaneously with the climatological minimum in the zonal background wind observed in the underlying column.

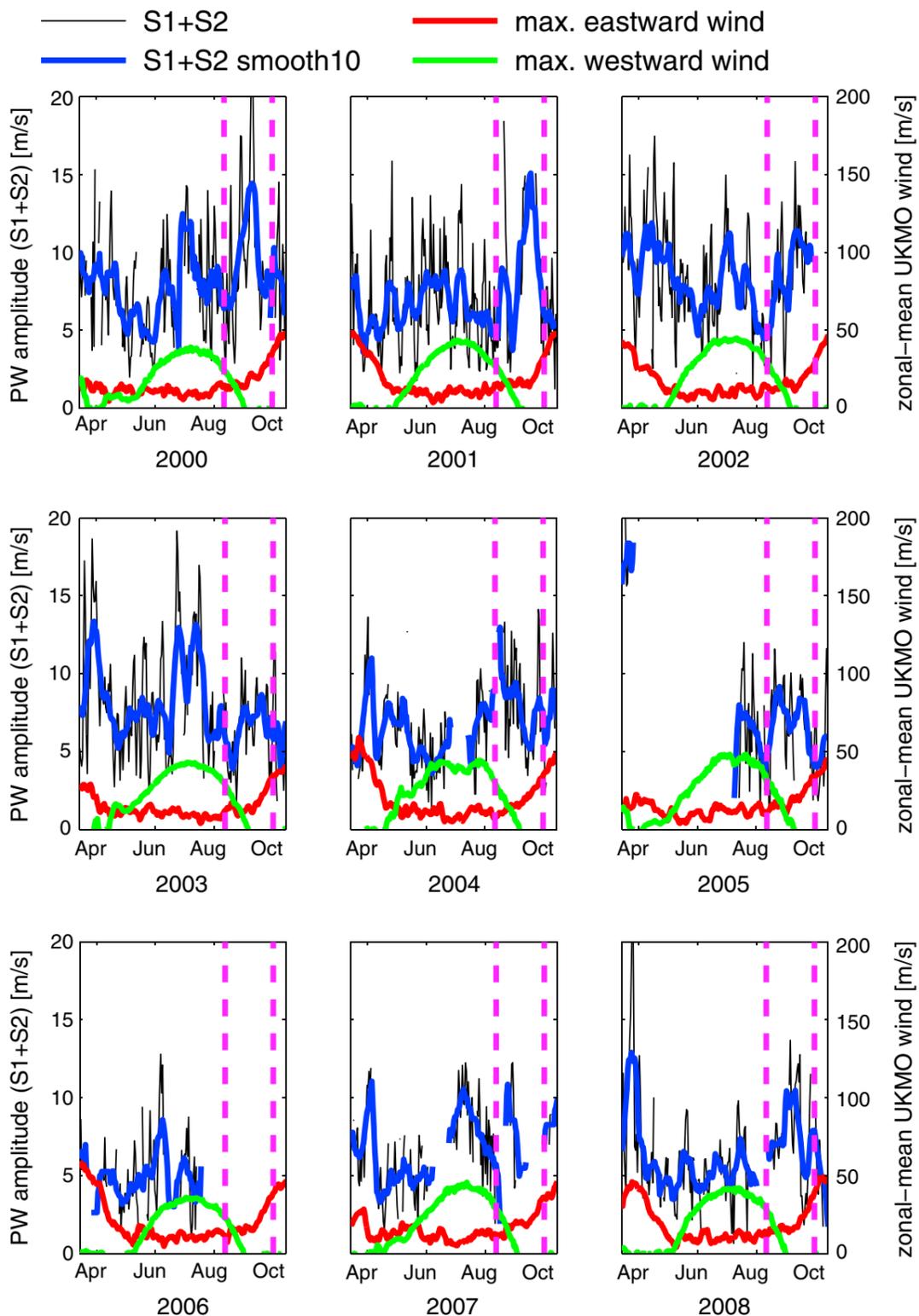

**Figure 2.** Ten day smoothed (blue, left axis) and unsmoothed (black, left axis) mesospheric (mean altitude around 95 km) PW activity ($S_1+S_2$) and maximum UKMO zonal mean eastward (red, right axis) and westward (green, right axis) wind below (∼65 km) between 1 March and 15 October for years 2000–2008. The vertical dashed lines mark the start and end of the PW enhancement seen in Figure 3.

Coincident with this burst of PW activity, a temporary poleward wind perturbation can be seen in the daily mean meridional wind of individual SuperDARN radars. As an example, Figure 4 shows the climatology of the daily mean meridional wind (blue) in the mesosphere (∼ 95 km) measured by the SuperDARN radar at Stokkseyri (64.7°N, 26.9°W) and the 20 day smoothing of the climatology (green). The strength and direction of the gravity wave forcing is related to the net filtering below the MLT and hence the difference between the maximum eastward and westward wind in the height column below the MLT. Fourier decomposition of this difference reveals only an annual and semiannual component (not shown). The expected meridional wind in the mesosphere can therefore be constructed by a fit of a bias, an annual, and a semiannual component. This expected wind is shown as the red line in Figure 4. There are clear perturbations in the measured mean wind (difference between green and red line) that cannot be explained by the effect of gravity wave forcing and radiation alone. During autumn a poleward meridional wind perturbation is seen to occur just prior to day 250 coincident with the enhancement in PW activity shown in Figures 2 and 3.

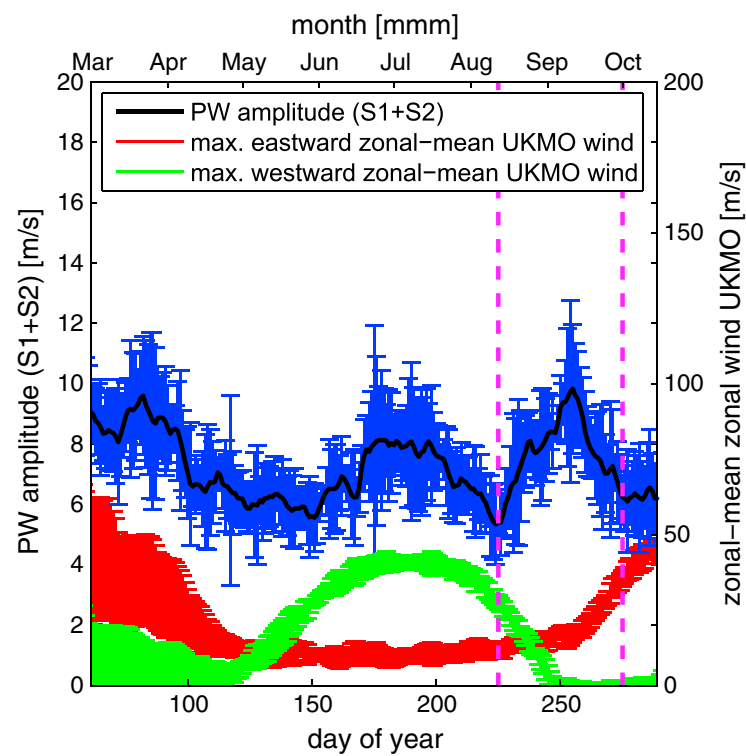

**Figure 3.** Ten day smoothed climatology (2000–2008) of mesospheric (mean altitude around 95 km) PW activity (black, left axis) with error bars of $1\sigma$ (blue) and climatology (2000–2008) of maximum eastward (red) and westward (green) zonal mean UKMO wind below ~65 km (right axis) between 1 March and 15 October.

## 4. Discussion

The vertical propagation of a PW depends on its phase speed relative to the zonal background wind in the atmosphere [e.g., *Salby*, 1996]. The maximum wind velocity in the stratosphere therefore gives an indication as to when and which waves can propagate through the stratosphere into the mesosphere. *Charney and Drazin* [1961] showed for a simplified analytically solvable system that PWs can only propagate into regions where the zonal mean wind ($\bar{u}$) is both larger (more eastward) than the zonal wave velocity ($c$) of the wave and smaller (more westward) than the Rossby critical velocity ($U_c$), where $U_c$ is dependent on the horizontal scale of the wave. PW activity in the middle atmosphere is generally observed to be weaker in summer than in winter [e.g., *Alexander and Shepherd*, 2010; *McDonald et al.*, 2011; *Day and Mitchell*, 2010], since the intrinsic PW velocity is always westward and their propagation is therefore strongly reduced by the summertime westward jets in the stratosphere. However, during the fall equinox, stratospheric zonal winds are generally low, resulting in a high atmospheric transmission and the enhancement of PW propagation into the MLT seen in Figures 2 and 3.

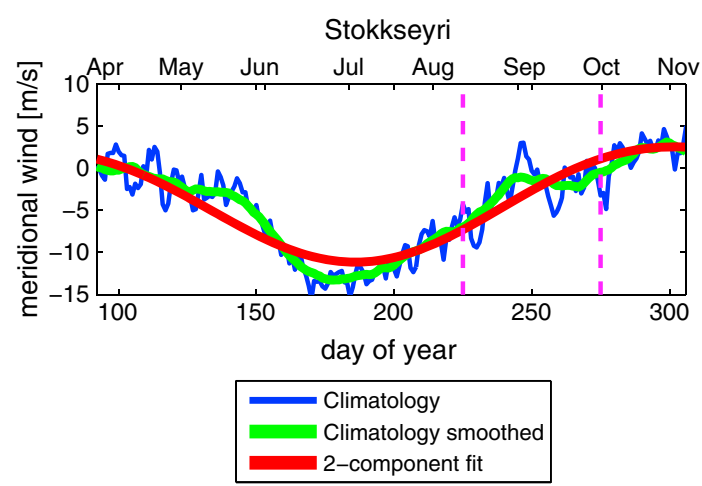

**Figure 4.** Climatology of planetary wave activity (blue) and 20 day smoothed climatology (green) from the Stokkseyri (64.7°N, 26.9°W) SuperDARN radar. The red curve depicts the expected seasonal variation of the meridional wind due to pure radiative and gravity wave forcing. A poleward perturbation away from the standard picture is clearly visible around fall equinox.

Bursts of 5 day and 16 day PW activity during the fall equinox have previously been observed in mesospheric winds derived from single northern hemispheric meteor radars at, e.g., Sheffield (53°N, 4°W), Esrange (68°N, 21°E), and Bear Lake (42°N, 11°W) by *Mitchell et al.* [1999], *Day and Mitchell* [2010], and *Day et al.* [2012], respectively. *Day and Mitchell* [2010] present the seasonal behavior of the quasi 5 day (4–7 days) wave in the MLT showing a clear enhancement of its amplitude in fall. This enhancement of a single temporal component occurs at the same time as the enhancement of the zonal wave numbers 1 and 2 PW structures observed here (Figure 3). Indeed, spectral analysis of the PW observed by the SuperDARN chain around fall equinox reveals quasi 5, 10, and 16 day oscillations in agreement with other PW observations at single stations around the fall equinox [e.g., *Day and Mitchell*, 2010; *Day et al.*, 2012]. In addition, Hovmöller diagrams (Figure 1 and *Kleinknecht et al.*, 2014, Figure 5) of the PW structures show that the wave-field during fall (August–September) often exhibits a strong phase shift and westward traveling wave components with periods shorter than 20 days become dominant in both zonal wave numbers. The strong wave transition toward the west at the time of the wave enhancement agrees with model simulations [*Liu et al.*, 2001].

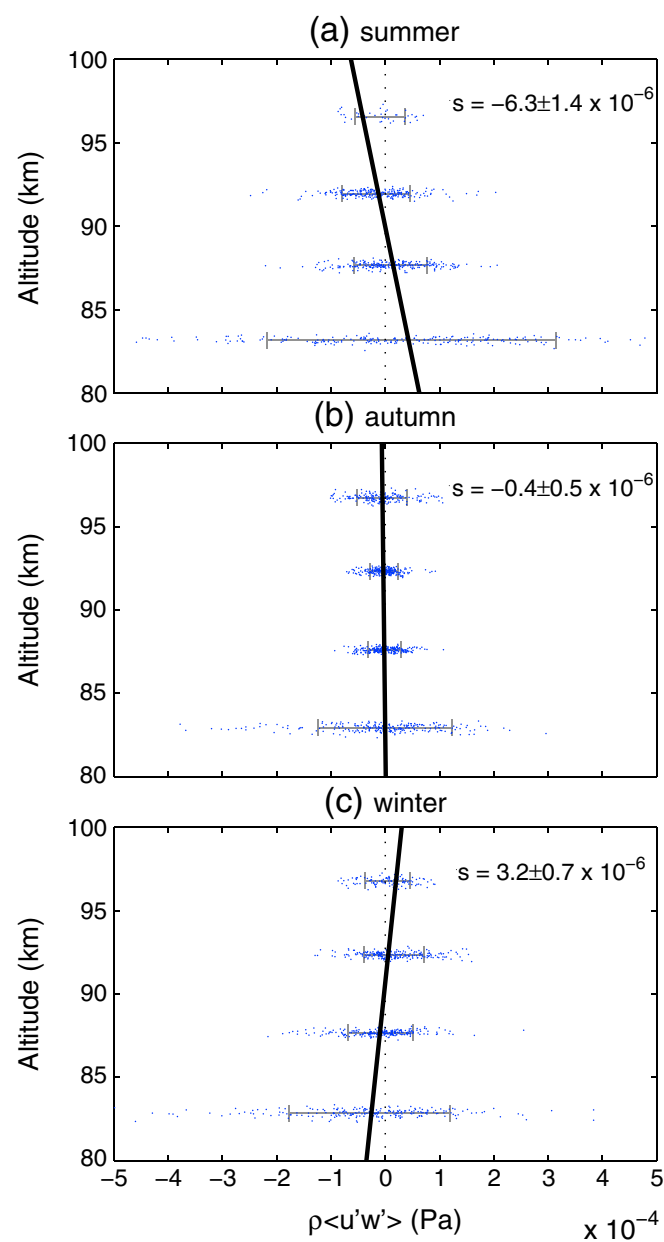

**Figure 5.** Vertical profile of hourly values of $\overline{\rho u'w'}$ (blue dots) and their $\pm 1\sigma$ standard deviation (grey line) together with a linear best fit (black, slope $s\left[\frac{km}{Pa}\right] \pm 1\sigma$ shown in figure) through all available data points for (a) days 156–166, (b) days 252–262, and (c) days 334–344 in the year 2013.

*Riggin et al.* [2003] observed an autumn equinox enhancement of semidiurnal tidal amplitudes from MF radar observations at these latitudes which was localized between 82 and 90 km at the preferred altitudes for these radars. The amplitude decay above this altitude was attributed to refraction of the tide by the background wind field. However, *Mitchell et al.* [2002], using a SKiYMET meteor radar system located at Esrange (68°N, 21°E), observed the amplitude of the autumn equinox semidiurnal tide to continue to grow up to 97 km altitude. Although refraction of the wave energy would be expected to produce a westward and poleward forcing of the background winds, the tendency for MF radars to progressively underestimate the winds and tidal amplitudes above 90 km is well documented [e.g., *Manson et al.*, 2004; *Portnyagin et al.*, 2004]. So although tidal forcing of the background wind may play a role during the autumn equinox, the SuperDARN radars lack the altitude resolution to quantify its contribution.

Upward propagation of gravity waves is also prohibited if the background wind matches the phase velocity of the wave [*Fritts and Alexander*, 2003]. However, since GWs can have both eastward and westward phase velocities, the seasonal change of the stratospheric background wind leads to the transmission of westward GWs in winter and, correspondingly, eastward waves in summer. Those GWs that reach the MLT and break deposit their momentum. The resulting drag creates a meridional flow that redirects the wind away from the summer and toward the winter pole [*Lindzen*, 1981; *Holton*, 1982, 1983; *Garcia and Solomon*, 1985]. During equinox the zonal winds are generally low, and therefore, the filtering is at a minimum for gravity waves. Hence, GW forcing in the MLT would be expected to minimize around equinox as the GW forcing makes a smooth transition from positive (summer) to negative (winter) forcing [*Andrews et al.*, 1987].

The net effect of GWs on the zonal momentum budget of the mesopause region can be derived from the vertical divergence of the density-weighted vertical flux of horizontal momentum [e.g., *Fritts and Vincent*, 1987]. To illustrate how the GW forcing characteristically varies between solstice and equinox conditions, vertical profiles of density-weighted GW fluxes of zonal momentum, $\rho \overline{u'w'}$, determined using the meteor radar located in Trondheim [*Hocking*, 2005; *de Wit et al.*, 2014] are presented in Figure 5. Figure 5b shows $\rho \overline{u'w'}$ (blue dots) over a period of 10 days around fall equinox (days 252–262) in 2013 and their $1\sigma$ standard deviation (horizontal grey lines) at each altitude bin, together with a linear best fit through all available data points representing the vertical profile of $\rho \overline{u'w'}$ (black line). For comparison, vertical profiles of $\rho \overline{u'w'}$ representative of summertime (days 156–166 2013) and wintertime (days 334–344 2013) forcing conditions are shown in Figures 5a and 5c, respectively. During summer (winter) $\rho \overline{u'w'}$ decreases (increases) significantly with height, indicating an eastward (westward) GW forcing. However, around the fall equinox, the vertical divergence of $\rho \overline{u'w'}$ is not statistically significantly different from zero, indicating no net zonal GW forcing of the MLT region. Consequently, in the absence of other MLT dynamic forcing, the atmosphere should relax toward radiative equilibrium around equinox with MLT meridional winds close to zero [*Andrews*, 2010].

In contrast, meridional wind observations in the MLT show a poleward directed perturbation during equinox, as can be seen in Figure 4. Such poleward bursts of MLT meridional winds during the fall equinox have previously been reported at similar northern and southern latitudes [e.g., *Sandford et al.*, 2010] but not related to specific forcing mechanisms. In addition, mesosphere temperature enhancements near equinox have been observed at a variety of longitudes in both hemispheres [e.g., *Taylor et al.*, 2001; *Manson et al.*, 2002; *Shepherd et al.*, 2002; *Espy and Stegman*, 2002; *French and Burns*, 2004]. Both the enhanced poleward flow and the temperature enhancement are evidence that the MLT is driven away from radiative equilibrium by westward wave forcing during the observed enhancement of PW amplitudes at the fall equinox, when the net GW forcing minimizes. It appears that the westward momentum carried by PWs temporarily dominates the forcing of the meridional wind, driving it toward the pole and enhancing temperatures in the mesosphere during the fall equinox due to convergence and subsequent downwelling.

It should be noted that the observed enhancement of PW amplitude seems to occur slightly earlier than climatological OH-temperature perturbations that have been observed [*Taylor et al.*, 2001; *Espy and Stegman*, 2002; *French and Burns*, 2004]. However, LIDAR temperature observations [*She et al.*, 2000; *Pan and Gardner*, 2003; *Kawahara et al.*, 2004] tend to show that the fall temperature maximum occurs first at the higher altitudes of the SuperDARN wind observations (95 km) and progresses downward in time to where the OH-temperature observations take place (87 km). This downward phase progression of the temperature perturbation is again indicative of a wave-breaking source [*Plumb and Semeniuk*, 2003].

Together, these observations indicate that the temporary increase in the westward momentum carried by the PWs temporarily drives the ageostrophic meridional circulation when GW forcing is weak and in a transition from positive (summer) to negative (winter) forcing during the fall equinox. This results in a temporary increase in the poleward flow, increasing mesospheric temperatures beyond that which the weak GW forcing alone would produce. As the stratospheric winds become steadily more eastward, the increasingly filtered GWs again assume control of the meridional residual circulation and the mesospheric temperature. Although a similar process occurs during spring equinox, its timing in the Northern Hemisphere is more variable due to the timing of the final warming of the stratosphere. Hence, the climatological mean of the springtime temperature enhancement in the Northern Hemisphere [*Espy and Stegman*, 2002; *Manson et al.*, 2002; *Shepherd et al.*, 2002] is less pronounced. Thus, the enhanced flux of westward momentum carried by PWs temporarily provides the main forcing of the MLT residual circulation, resulting in enhanced poleward flow and the adiabatic heating observed in OH airglow and LIDAR temperature measurements in the MLT during fall equinox.

## 5. Conclusion

Evidence for temporary PW forcing of the MLT during the fall equinox has been presented and discussed as a possible cause of a concurrent poleward wind perturbation and subsequent temperature enhancement in the MLT around fall equinox. PW activity of zonal wave numbers 1 and 2 has been extracted using a longitudinal chain of high-latitude Northern Hemisphere SuperDARN radars over 9 years (2000–2008). A clear and significant enhancement of the PW amplitudes in the MLT is observed each year, as well as in the climatological composite, during the fall equinox. At this time the low zonal wind speeds in the underlying atmosphere provide reduced filtering of the GWs and PWs. While the resulting enhanced transmission of PWs increases the westward momentum carried to the MLT, the net momentum deposition of the more isotropic GWs is observed to decrease due to the reduced filtering. This temporary increase of westward momentum transmitted to the MLT by the PWs then controls the meridional residual circulation, resulting in the increased poleward flows observed here and by other studies. This, in turn, raises mesospheric temperatures beyond that which the weakened momentum flux provided by the GWs would provide and results in the enhanced mesospheric temperatures that have been observed previously. In the future the proposed mechanism to characterize the role of PW forcing on the meridional circulation will be tested by appropriate modeling efforts.

**Acknowledgments**
The authors acknowledge the use of SuperDARN data. SuperDARN is a collection of radars funded by the national scientific funding agencies of Australia, Canada, China, France, Japan, South Africa, United Kingdom, and United States of America. The UK Meteorological Office assimilated data set was made available by the British Atmospheric Data Centre. This study was partly supported by the Research Council of Norway/CoE under contract 223252/F50.

The Editor thanks two anonymous reviewers for their assistance in evaluating this paper.